\documentclass[10pt,a4paper]{article}
\usepackage{xcolor}
\usepackage{mdframed}
\usepackage[utf8]{inputenc}
\usepackage{amsmath, amsthm, amssymb, bbm, bm}
\usepackage{tikz-cd}
\usepackage{enumerate}
\usepackage{appendix}
\usepackage[hidelinks]{hyperref}
\usepackage{color}

\definecolor{mygray}{gray}{0.75}

\newtheorem{tvrz}{Proposition}[section]

\newtheorem{theorem}[tvrz]{Theorem}
\newtheorem{cor}[tvrz]{Corollary}
\theoremstyle{definition}

\theoremstyle{remark}

\theoremstyle{definition}
\newtheorem{mdexample}[tvrz]{Example}
{\begin{mdframed}[topline=false, rightline=false, bottomline=false, linewidth=0.2em, linecolor=mygray, innerleftmargin=0.5em, innerrightmargin=0,leftmargin=-0.7em]\begin{mdexample}}%
{\end{mdexample}\end{mdframed}}

\def\^{\wedge}
\def\<{\langle}
\def\>{\rangle}
\def\dal{\langle \! \langle}
\def\dar{\rangle \! \rangle}
\def\^{\wedge}
\def\cD{\nabla}
\def\g{\mathfrak{g}}
\def\d{\mathfrak{d}}
\def\RS{\mathcal{R}}
\def\A{\mathcal{A}}
\def\cN{\mathcal{N}}
\def\X{\mathfrak{X}}
\def\R{\mathbb{R}}
\def\D{\mathcal{D}}
\def\V{\mathcal{V}}
\def\K{\mathcal{K}}
\def\dr{\mathrm{d}}

\def\1{\mathbbm{1}}
\def\gm{\mathbf{G}}
\def\fK{\mathbf{K}}
\def\gTM{\mathbb{T}M}
\def\dM{\mathsf{M}}
\def\fvarphi{\boldsymbol{\varphi}}

\newcommand{\Li}[1]{\mathcal{L}_{#1}}

\DeclareMathOperator{\Tr}{Tr}
\DeclareMathOperator{\rk}{rk}
\DeclareMathOperator{\gr}{gr}
\DeclareMathOperator{\Ric}{Ric}
\DeclareMathOperator{\Div}{div}
\DeclareMathOperator{\LC}{LC}

\begin{document}
\begin{flushright}
\today
% preprint number (if any)
\end{flushright}
\vspace{0.7cm}
\begin{center}
 %\vskip1cm

\baselineskip=13pt {\Large \bf{Palatini Variation in Generalized Geometry and String Effective Actions}\\}
 \vskip0.5cm
 \vskip0.7cm
 {\large{Branislav Jurčo$^{1}$, Filip Moučka$^{2}$, Jan Vysoký$^{2}$}}\\
 \vskip0.6cm
$^{1}$\textit{Charles University Prague, 
Faculty of Mathematics and Physics, 
Mathematical Institute\\ Sokolovská 49/83, 186 75 Prague 8, Czech Republic}\\ 
$^{2}$\textit{Faculty of Nuclear Sciences and Physical Engineering, Czech Technical University in Prague\\ Břehová 7, 115 19 Prague 1, Czech Republic, }\\
\vskip0.3cm
\end{center}

\begin{abstract}
We develop the Palatini formalism within the framework of generalized Riemannian geometry of Courant algebroids. In this context, the Palatini variation of a generalized Einstein--Hilbert--Palatini action - formed using a generalized metric, a Courant algebroid connection (in contrary to the ordinary case, not necessarily a torsionless one) and a volume form - leads naturally to a proper notion of a generalized Levi-Civita connection and low-energy effective actions of string theory. 
\end{abstract}

\section{Introduction} \label{sec_intro}
In general relativity, one describes spacetimes, pairs $(M, g)$ with  $M$ being a (pseudo-)Riemannian manifold with a metric $g$. The field equations are obtained by metric variation of the action functional (Einstein--Hilbert action), the spacetime integral of the scalar curvature formed using the Levi-Civita connection (Christoffel symbols). The Levi-Civita connection is uniquely determined requiring zero torsion and metric compatibility, and is expressed explicitly using only the metric.
 
In an alternative approach known under the name Palatini variation, the action is also a spacetime integral of the scalar curvature. This time, however, one considers a general torsion-free affine connection on $M$. There is a priori no relation between the connection and the metric, these are independent fields of the theory and the equations of motion are obtained from variation with respect to these. The compatibility of the connection with a metric is now a consequence of equations of motion. Hence, the action used in the Palatini formulation and the Einstein--Hilbert action are classically equivalent. See e.g. \cite{ferraris1982variational} for a detailed historical overview and references. 

In the context of generalized Riemannian geometry, ordinary affine connections can naturally be generalized to what is called Courant algebroid connections \cite{alekseevxu}, \cite{2007arXiv0710.2719G}. Generalization of the torsion \cite{alekseevxu},\cite{2007arXiv0710.2719G} and curvature \cite{Jurco:2016emw}, cf. for closely related definition in context of double field theory \cite{Hohm:2012mf}, is more tricky. In contrary to the ordinary case, there is no obvious geometric meaning to these. Also, in general, there are infinitely many Levi-Civita connections. Even the set of Levi-Civita  connections with a prescribed divergence will, in general, still remain to be infinite. The most natural choice for the divergence fixing, is the one associated with a volume form on the spacetime. This is how the dilaton field can be related to the freedom in the definition of Levi-Civita property. Nevertheless, one can still form a well defined Einstein--Hilbert action, using the generalized curvature formed from a Levi-Civita connection with a divergence fixed using the volume form. This action depends only on the Courant algebroid, generalized metric on it and the volume form, which in the simplest case are given by a Riemannian metric $g$, Kalb-Ramond field $B$ and the dilaton $\phi$. It can be shown that, depending on the choice of the Courant algebroid, this action reproduces the NS-NS sector of the respective string low-energy effective action \cite{Coimbra:2011nw,Jurco:2016emw,Severa:2018pag}. The equations of motion do not depend on the choice of Levi-Civita connection.

In the present paper, we extend the Palatini method to the setting of generalized Riemannian geometry of Courant algebroids, cf. \cite{Jurco:2016emw} and references therein. Hence, the independent fields will be a Courant algebroid connection, generalized metric and a volume form. In analogy with the ordinary case, the Levi-Civita property and now also the divergence fixing are consequences of the equations of motion. Remarkably, we do not have to assume that the connection is torsion-free. In particular, this justifies the generalized geometry notion of the torsion, Levi-Civita property and elucidates the role of the dilaton in the generalized geometry. Let us also note that the Palatini variation, as presented here, is compatible with reduction of Courant algebroids \cite{Bursztyn2007DiracGQ}, i.e. generalized geometry version of Kaluza--Klein reduction \cite{Vysoky:2017epf}, and hence with Poisson-Lie T-duality originating from such reductions \cite{Severa:2015hta}.

The paper is organized as follows. 

In Section \ref{sec_math}, we quickly introduce all mathematical tools required in this paper. Since there is already a plethora of literature on those, we omit details. We recall the notion of a Courant algebroid, generalized metric and Courant algebroid connection. 

Section \ref{sec_palatini} contains the main statement of this paper. It shows that a generalized metric, volume form and a Courant algebroid connection can be used to form a simple and elegant action functional which we call the generalized Palatini action. We state the theorem describing its equations of motion. Notably, this bears a striking resemblance to the famous Palatini variation in general relativity: the equation of motion obtained by the variation with respect to the connection implies its compatibility with the remaining two fields of the action, the generalized metric and the volume form. Interestingly, in our generalized setting, one also automatically obtains the torsion-freeness condition. 

It is argued that the equation of motion for the connection can be solved and the solution used to obtain a physically relevant Einstein--Hilbert action. 

We list several examples to justify the physical relevance of this paper in Section \ref{sec_examples}. For exact Courant algebroids, we obtain a certain sector of type II supergravity. A particular class of transitive Courant algebroids, called heterotic ones, leads to the geometrical description of heterotic supergravity. A proper choice of  Dirac structure and of its complement allows one to find an equivalent formulation of type II supergravity, called symplectic gravity. Finally, we show how situation simplifies for a Courant algebroid over a point - a quadratic Lie algebra. 

In Section \ref{sec_proof}, we give the proof of the main theorem. 

\subsection{Acknowledgements}
The research of B.J. was supported by grant GA\v CR EXPRO 19–28628X.
J.V. is grateful for financial support from MŠMT under grant no. RVO 14000. F.M. thanks for the financial support to the Grant Agency of the Czech Technical University in Prague, grant No. SGS22/178/OHK4/3T/14.

This paper gives an improved and extended account of results announced by the F.M. in \cite{Filip}. 
\section{Preliminaries} \label{sec_math}
In this section, we shall recall basic mathematical notions used in this paper. Let us give only a bare minimum of definitions and basic properties. Most of the details can be found, e.g., in \cite{Jurco:2016emw} and in references cited therein. Throughout the paper, $M$ will be a fixed orientable ($n$-dimensional, smooth) manifold.

\begin{enumerate}[(1)]
\item A \textbf{Courant algebroid} $(E,\rho,\<\cdot,\cdot\>_{E},[\cdot,\cdot]_{E})$ consists of the following objects:
\begin{enumerate}[(i)]
\item $E$ is a vector bundle over $M$. $\Gamma(E)$ denotes the module of its smooth sections.
\item $\rho: E \rightarrow TM$ is a vector bundle map called \textbf{the anchor}.
\item $\<\cdot,\cdot\>_{E}$ is a fiber-wise metric on $E$. We will also sometimes write $\mathbf{g}_{E}$ for $\<\cdot,\cdot\>_{E}$. 
\item $[\cdot,\cdot]_{E}$ is an $\R$-bilinear bracket on $\Gamma(E)$.
\end{enumerate}
These are subject to the set of following axioms:
\begin{enumerate}[({a}1)]
\item $[\psi,f\psi']_{E} = f [\psi,\psi']_{E} + (\rho(\psi)f)\psi'$, 
\item $\rho(\psi)\<\psi',\psi''\>_{E} = \<[\psi,\psi']_{E},\psi''\>_{E} + \< \psi', [\psi,\psi'']_{E}\>_{E}$,
\item $[\psi,[\psi',\psi'']_{E}]_{E} = [[\psi,\psi']_{E},\psi'']_{E} + [\psi',[\psi,\psi'']_{E}]_{E}$,
\item $\< [\psi,\psi]_{E},\psi'\>_{E} = \frac{1}{2} \rho(\psi')\<\psi,\psi\>_{E}$,
\end{enumerate}
for all $\psi,\psi',\psi'' \in \Gamma(E)$ and $f \in C^{\infty}(M$).
\item Let $E$ be a vector bundle over $M$ equipped with a fiber-wise metric $\<\cdot,\cdot\>_{E}$. A \textbf{generalized metric} on $(E,\<\cdot,\cdot\>_{E})$ is a maximal positive subbundle $V_{+} \subseteq E$ with respect to $\<\cdot,\cdot\>_{E}$. Let us summarize some of its basic properties:
\begin{enumerate}[(i)]
\item There is an induced decomposition $E = V_{+} \oplus V_{-}$, where $V_{-} = V_{+}^{\perp}$ is the orthogonal complement taken with respect to $\<\cdot,\cdot\>_{E}$.
\item It induces an orthogonal vector bundle endomorphism $\tau: E \rightarrow E$ satisfying $\tau^{2} = \1_{E}$, such that $V_{\pm}$ are its $\pm 1$ eigenbundles. 
\item For each $\psi \in \Gamma(E)$, let $\psi_{\pm}$ denote its components in $\Gamma(V_{\pm})$. Then the formula 
\begin{equation} \label{eq_gmetric}
\gm(\psi,\psi') := \<\psi_{+},\psi'_{+}\>_{E} - \<\psi_{-},\psi'_{-}\>_{E}
\end{equation}
defines a positive definite fiber-wise metric $\gm$ on $E$. Equivalently, one can define it by $\gm(\psi,\psi') = \<\psi, \tau(\psi')\>_{E}$ for all $\psi,\psi' \in \Gamma(E)$. 
\end{enumerate}
\item Let $(E,\rho,\<\cdot,\cdot\>_{E},[\cdot,\cdot]_{E})$ be a Courant algebroid. A \textbf{Courant algebroid connection}\footnote{\cite{alekseevxu}, \cite{2007arXiv0710.2719G}} on $E$ is an $\R$-bilinear map $\cD: \Gamma(E) \times \Gamma(E) \rightarrow \Gamma(E)$ satisfying
\begin{enumerate}[(b1)]
\item $\cD(f \psi, \psi') = f \cD(\psi,\psi')$, 
\item $\cD(\psi, f\psi') = f \cD(\psi,\psi') + (\rho(\psi)f) \psi'$,
\item $\rho(\psi)\<\psi',\psi''\>_{E} = \< \cD(\psi,\psi'),\psi''\>_{E} + \< \psi', \cD(\psi,\psi'')\>_{E}$,
\end{enumerate}
for all $\psi,\psi',\psi'' \in \Gamma(E)$ and $f \in C^{\infty}(M)$. The operator $\cD_{\psi} := \cD(\psi,\cdot)$ is called a \textbf{covariant derivative} along $\psi \in \Gamma(E)$. Every Courant algebroid connection $\cD$ induces several additional structures on $E$:
\begin{enumerate}[(i)]
\item There is a \textbf{torsion $3$-form}\footnote{\cite{alekseevxu}, \cite{2007arXiv0710.2719G}} $T_{\cD} \in \Gamma(\Lambda^{3}E^{\ast})$ defined as 
\begin{equation}
T_{\cD}(\psi,\psi',\psi'') := \< \cD_{\psi}\psi' - \cD_{\psi'}\psi - [\psi,\psi']_{E},\psi''\>_{E} + \< \cD_{\psi''}\psi,\psi'\>_{E},
\end{equation}
for all $\psi,\psi',\psi'' \in \Gamma(E)$. We say that $\cD$ is \textbf{torsion-free}, if $T_{\cD} = 0$. 
\item A \textbf{divergence} on $E$ is any $\R$-linear map $\Div: \Gamma(E) \rightarrow C^{\infty}(M)$ satisfying
\begin{equation}
\Div(f\psi) = f \Div(\psi) + \rho(\psi)f,
\end{equation}
for all $\psi \in \Gamma(E)$ and $f \in C^{\infty}(M)$. To any Courant algebroid connection $\cD$, there is an associated divergence $\Div_{\cD}$ defined, for each $\psi \in \Gamma(E)$, as 
\begin{equation}
\Div_{\cD}(\psi) := \Tr( \cD(\cdot,\psi)).
\end{equation}
Another divergence $\Div_{\omega}: \Gamma(E) \rightarrow C^{\infty}(M)$ can be constructed using a volume form $\omega \in \Omega^{n}(M)$ on the manifold $M$. It is defined, for each $\psi \in \Gamma(E)$, as 
\begin{equation} \label{eq_divomega}
\Div_{\omega}(\psi) := \Li{\rho(\psi)}(\omega) \cdot \omega^{-1}. 
\end{equation}
\item There is a \textbf{curvature tensor}\footnote{Compare with a related definition in double field theory \cite{Hohm:2012mf}.} $R_{\cD}$ defined as follows. First, define
\begin{equation}
R^{0}_{\cD}(\phi',\phi,\psi,\psi') := \< \cD_{\psi}(\cD_{\psi'}\phi) - \cD_{\psi'}(\cD_{\psi}\phi) - \cD_{[\psi,\psi']_{E}}\phi, \phi'\>_{E},
\end{equation}
for all $\psi,\psi',\phi,\phi' \in \Gamma(E)$. This is not a tensor on $E$, nor it has any nice symmetries. The actual definition then reads\footnote{Let us notice that in \cite{2020JHEP...01..007B} an alternative but equivalent definitions of torsion and curvature, which look formally as the ordinary ones, are given for the exact Courant algebroid at the expense of modifying the Courant bracket.}
\begin{equation}
R_{\cD}(\phi',\phi,\psi,\psi') := \frac{1}{2} \{ R^{0}_{\cD}(\phi',\phi,\psi,\psi') + R^{0}_{\cD}(\psi',\psi,\phi,\phi') + \< \fK(\psi,\psi'), \fK(\phi,\phi')\>_{E}\}, 
\end{equation}
for all $\psi,\psi',\phi,\phi' \in \Gamma(E)$, where $\< \fK(\psi,\psi'),\phi \>_{E} := \< \cD_{\phi}\psi,\psi'\>_{E}$. It can be shown that $R_{\cD}$ is $C^{\infty}(M)$-linear in every input and it enjoys the following symmetries:
\begin{align}
R_{\cD}(\phi',\phi,\psi,\psi') + R_{\cD}(\phi',\phi,\psi',\psi) = & \ 0, \\
R_{\cD}(\phi',\phi,\psi,\psi') + R_{\cD}(\phi,\phi',\psi,\psi') = & \ 0, \\
R_{\cD}(\phi',\phi,\psi,\psi') - R_{\cD}(\psi',\psi,\phi,\phi') = & \ 0, 
\end{align}
for all $\psi,\psi',\phi,\phi' \in \Gamma(E)$. These symmetries allow for an unambiguous definition of the \textbf{Ricci tensor} $\Ric_{\cD}$, defined for all $\psi,\psi' \in \Gamma(E)$ as a partial trace
\begin{equation}
\Ric_{\cD}(\psi,\psi') = \Tr_{\mathbf{g}_{E}}( R_{\cD}(\cdot,\psi,\cdot,\psi')).
\end{equation}
It follows that $\Ric_{\cD}$ is symmetric in its inputs. Finally, for any fiber-wise metric $\mathbf{g}$ on $E$, one can define the corresponding \textbf{scalar curvature} as 
\begin{equation}
\RS_{\cD}^{\mathbf{g}} := \Tr_{\mathbf{g}}( \Ric_{\cD}). 
\end{equation}
\item Let $V_{+}$ be a generalized metric. We say that $\cD$ is \textbf{metric compatible with $V_{+}$}, if 
\begin{equation}
\cD_{\psi}( \Gamma(V_{+})) \subseteq \Gamma(V_{+}),
\end{equation}
for all $\psi \in \Gamma(E)$. Equivalently, using the fiber-wise metric (\ref{eq_gmetric}), this can be written as 
\begin{equation}
\rho(\psi)\gm(\psi',\psi'') = \gm( \cD_{\psi}\psi',\psi'') + \gm(\psi', \cD_{\psi}\psi''),
\end{equation}
for all $\psi,\psi',\psi'' \in \Gamma(E)$. This can be compared with axiom (b3) for $\cD$. 
\end{enumerate}
\item Let $(E,\rho,\<\cdot,\cdot\>_{E},[\cdot,\cdot]_{E})$ be a Courant algebroid equipped with a generalized metric $V_{+}$ and a Courant algebroid connection $\cD$. Let $(p,q)$ denote the signature of $\<\cdot,\cdot\>_{E}$. 
\begin{enumerate}[(i)]
\item $\cD$ is called \textbf{Levi-Civita with respect to $V_{+}$}, if it is metric compatible with $V_{+}$ and torsion-free. We write $\cD \in \LC(E,V_{+})$. It can be shown that $\LC(E,V_{+}) \neq \emptyset$. Moreover, unless both $p \in \{0,1\}$ and $q \in \{0,1\}$, this set is infinite. In general, there is thus no Koszul formula.
\item Let $\Div: \Gamma(E) \rightarrow C^{\infty}(M)$ be an arbitrarily fixed divergence. If $\cD$ is Levi-Civita with respect to $V_{+}$ and $\Div_{\cD} = \Div$, we write $\cD \in \LC(E,V_{+},\Div)$.  For $p,q \neq 1$, one can show that $\LC(E,V_{+},\Div) \neq \emptyset$. Note that also this set is infinite. 
\item $\cD$ is called \textbf{Ricci compatible with $V_{+}$}, if the Ricci tensor $\Ric_{\cD}$ is block-diagonal with respect to the decomposition $E = V_{+} \oplus V_{-}$, that is $\Ric_{\cD}(\Gamma(V_{+}),\Gamma(V_{-})) = 0$. 
\item Let $\cD \in \LC(E,V_{+},\Div_{\omega})$ for some volume form $\omega$ on $M$, see (\ref{eq_divomega}). It can be shown that the scalar curvature $\RS_{\cD}^{\mathbf{g}_{E}}$ is actually independent on $V_{+}$, $\cD$ and $\omega$. Consequently, we will denote this canonical scalar function associated to any Courant algebroid as $\RS_{E}$. 
\end{enumerate}
\end{enumerate}

\section{Palatini variation} \label{sec_palatini}
For \textit{any} Courant algebroid $(E,\rho,\<\cdot,\cdot\>_{E},[\cdot,\cdot]_{E})$, let us consider the action functional 
\begin{equation} \label{eq_action}
S_{P}[V_{+},\omega,\cD] := \int_{M} \{ \Tr_{\gm}(\Ric_{\cD}) + \RS_{E} \} \cdot \omega
\end{equation}
where the dynamical fields are the following:
\begin{enumerate}[(i)]
\item $V_{+}$ is a generalized metric on $(E,\<\cdot,\cdot\>_{E})$ inducing the fiber-wise metric $\gm$ via (\ref{eq_gmetric});
\item $\omega$ is an arbitrary volume form on $M$;
\item $\cD$ is \textit{any} Courant algebroid connection on $E$.
\end{enumerate}
Let us call $S_{P}$ a \textbf{generalized Palatini action}.

Note that the Ricci tensor $\Ric_{\cD}$ is defined using only $\cD$ and the data of the Courant algebroid $(E,\rho,\<\cdot,\cdot\>_{E},[\cdot,\cdot]_{E})$. One only uses the generalized metric $V_{+}$ to obtain the scalar curvature $\RS_{\cD}^{\gm} := \Tr_{\gm}(\Ric_{\cD})$. The resulting scalar function is then summed with the canonical scalar function $\RS_{E}$ and integrated over $M$ using the volume form $\omega$. All three dynamical fields are a priori not related in any way.  

The pivotal result of this paper concerns the conditions of extremality of this action (equations of motion). We formulate them in the form of a theorem. Let $(p,q)$ denote the signature of $\<\cdot,\cdot\>_{E}$. In everything what follows, we assume $p,q \neq 1$. 
\begin{theorem}[\textbf{Palatini variation}] \label{thm_palatini}
The fields $(V_{+},\omega,\cD)$ extremalize the action (\ref{eq_action}) if and only if the following conditions are satisfied:
\begin{enumerate}[(i)]
\item $\cD$ is Ricci compatible with $V_{+}$.
\item The sum of scalar curvatures vanishes, $\RS^{\gm}_{\cD} + \RS_{E} = 0$, where $\RS^{\gm}_{\cD} \equiv \Tr_{\gm}(\Ric_{\cD})$. 
\item The connection $\cD$ has the following properties:
\begin{enumerate}[(a)]
\item It is Levi-Civita with respect to $V_{+}$. 
\item The divergence operator of $\cD$ for all $\psi \in \Gamma(E)$ takes the form
\begin{equation} \label{eq_divergence}
\Div_{\cD}(\psi) = \Div_{\omega}(\psi) \equiv \Li{\rho(\psi)}(\omega) \cdot \omega^{-1}. 
\end{equation}
\end{enumerate}
\end{enumerate}
\end{theorem}
We will prove the theorem in Section \ref{sec_proof}. 

There are some remarks in order. 

First, the conditions $(i)$-$(iii)$ come from the independent variations of the fields $(V_{+},\omega,\cD)$, in this order. Second, the combination of the condition $(a)$ and $(b)$ can be simply written  as 
\begin{equation}
\cD \in \LC(E,V_{+},\Div_{\omega}). 
\end{equation}
As discussed in Section \ref{sec_math}-(4)-(ii), this set is non-empty and the condition $(iii)$ can be thus always solved. As already noted, this does not determine $\cD$ uniquely. However, if we choose $\cD$ from the subset $\LC(E,V_{+},\Div_{\omega})$, both the off-diagonal components of $\Ric_{\cD}$ and the scalar curvature $\RS_{\cD}^{\gm}$ do not depend on this choice. 

This observation allows one to define a generalized \textbf{Einstein--Hilbert action}
\begin{equation} \label{eq_EHaction}
S_{EH}[V_{+},\omega] := \int_{M} \{ \RS_{\cD}^{\gm} + \RS_{E} \} \cdot \omega,
\end{equation}
where $(V_{+},\omega)$ are the same as for $S_{P}$, but $\cD$ is now assumed to be an arbitrary element of the set $\LC(E,V_{+},\Div_{\omega})$ and it is no longer a dynamical field. In other words, we have 
\begin{equation} 
S_{EH}[V_{+},\omega] = S_{P}[V_{+},\omega,\cD(V_{+},\omega)],
\end{equation}
where $\cD = \cD(V_{+},\omega)$ is an arbitrary solution of the condition $(iii)$ in Theorem \ref{thm_palatini}. We also immediately obtain its equations of motion.
\begin{cor} \label{cor_EHEOM}
The fields $(V_{+},\omega)$ extremalize the action (\ref{eq_EHaction}), if and only if the following conditions are satisfied:
\begin{enumerate}[(i)]
\item $\cD$ is Ricci compatible with $V_{+}$.
\item The sum of scalar curvatures vanishes, $\RS^{\gm}_{\cD} +\RS_{E}  = 0$.
\end{enumerate}
Note that unlike in Theorem \ref{thm_palatini}, both conditions are equations for the variables $(V_{+},\omega)$ only. 
\end{cor}
To conclude this section, let us remark that some of the technical assumptions can be relaxed. In particular, $V_{+}$ does not have to be a positive subbundle, it suffices to require $V_{+} \cap V_{-} = 0$ so that we still can write $E = V_{+} \oplus V_{-}$. We then only have to assume that $\rk(V_{\pm}) > 1$. 

Finally, note that the axioms (a1) and (a3) for $(E,\rho,\<\cdot,\cdot\>_{E},[\cdot,\cdot]_{E})$ imply the equation
\begin{equation} \label{eq_rhohom}
\rho([\psi,\psi']_{E}) = [\rho(\psi),\rho(\psi')],
\end{equation}
for all $\psi,\psi' \in \Gamma(E)$. It turns out that we do not have to assume the full Jacobi identity (a3) and it suffices to consider ``almost Courant algebroids'' where we only require (\ref{eq_rhohom}) to hold. 
\section{Examples} \label{sec_examples}
Let us now discuss some examples relevant for physics. We will not provide detailed calculations as they are already present in our previous papers. 

\begin{enumerate}[(1)]
\item \textbf{Type II supergravity}: Let us consider the standard exact Courant algebroid on $E = \gTM := TM \oplus T^{\ast}M$, where $[\cdot,\cdot]_{E}$ is an $H$-twisted Dorfman bracket for a given closed $H \in \Omega^{3}(M)$. 

In this case, every generalized metric $V_{+} \subseteq E$ corresponds uniquely to a pair $(g,B)$, where $g$ is a Riemannian metric on $M$ and $B \in \Omega^{2}(M)$. More precisely, one has 
\begin{equation}
\Gamma(V_{\pm}) = \{ (\pm g + B)(X) \; | \; X \in \X(M) \}.
\end{equation}
Next, assuming that $M$ is connected, every volume form $\omega$ on $M$ can be written as $\omega = \pm e^{-2\phi} \omega_{g}$ for a unique scalar function $\phi \in C^{\infty}(M)$. Here, $\omega_{g}$ denotes the metric volume form associated to $g$ (and a chosen orientation of $M$). The condition (\ref{eq_divergence}) now becomes
\begin{equation}
\Div_{\cD}(X,\xi) = \Div_{g}(X) - 2 (\dr \phi)(X),
\end{equation}
for all $(X,\xi) \in \Gamma(E)$, where $\Div_{g}$ is the usual divergence operator on vector fields associated to the metric $g$. Now, assuming that $\cD \in \LC(E,V_{+},\Div_{\omega})$, one can calculate the scalar curvature $\RS_{\cD}^{\gm}$, the canonical function $\RS_{E}$ and express them in terms of $(g,B,\phi)$. It turns out that 
\begin{equation}
\RS_{\cD}^{\gm} = \RS(g) - \frac{1}{2}\<H + \dr{B}, H + \dr{B}\>_{g} + 4 \Delta_{g}(\phi) - 4 \| \cD^{g} \phi \|^{2}_{g}, \; \; \RS_{E} = 0,
\end{equation}
where $\RS(g)$ is the usual scalar curvature of metric $g$, $\Delta_{g}$ is the Laplace-Bertrami operator and $\cD^{g} \phi \in \X(M)$ is a gradient of $\phi$. By plugging this into the generalized Einstein--Hilbert action, we obtain 
\begin{equation}
S_{EH}[g,B,\phi] = \int_{M} e^{-2 \phi} \{ \RS(g) - \frac{1}{2}\<H + \dr{B}, H + \dr{B}\>_{g} + 4 \Delta_{g}(\phi) - 4 \| \cD^{g} \phi \|^{2}_{g} \} \cdot \omega_{g},
\end{equation}
If we assume that $\partial M = \emptyset$ or $\dr{\phi}|_{\partial M} = 0$, we may get rid of $\Delta_{g}$ and write 
\begin{equation}
S_{EH}[g,B,\phi] = \int_{M} e^{-2 \phi} \{ \RS(g) - \frac{1}{2}\<H + \dr{B}, H + \dr{B}\>_{g} + 4 \| \cD^{g} \phi \|^{2}_{g} \} \cdot \omega_{g}. 
\end{equation}
But this is precisely the Neveu-Schwarz sector of the type II supergravities coming from superstring theory, see \cite{polchinski2005string}. Naturally, there are no restrictions on dimensions since we omit the Ramond-Ramond and Chern-Simons sectors completely. We have shown explicitly in \cite{Jurco:2016emw} that equations of motion of this action obtained directly by varying the fields $(g,B,\phi)$ indeed correspond to the conditions in Corollary \ref{cor_EHEOM}. 
\item \textbf{Heterotic supergravity}: Let $G$ be a compact Lie group with a Lie algebra $(\g,[\cdot,\cdot]_{\g})$. In particular, the corresponding Cartan-Killing form $\<\cdot,\cdot\>_{\g}$ is negative definite. Suppose $\pi: P \rightarrow M$ is a principal $G$-bundle over $M$ and fix a principal $G$-bundle connection $A \in \Omega^{1}(P,\g)$.  

This time, let $E = TM \oplus \g_{P} \oplus T^{\ast}M$, where $\g_{P}$ is the adjoint bundle associated to $P$. The fiber-wise metric $\<\cdot,\cdot\>_{E}$ takes the form
\begin{equation}
\< (X,\Phi,\xi), (Y,\Psi,\eta) \>_{E} = \eta(X) + \xi(Y) + \<\Phi,\Psi\>_{\g},
\end{equation}
for all $(X,\Phi,\xi),(Y,\Psi,\eta) \in \Gamma(E)$, where $\<\cdot,\cdot\>_{\g}$ denotes the fiber-wise metric on $\g_{P}$ induced by the Cartan-Killing form. The Courant algebroid bracket $[\cdot,\cdot]_{E}$ reads
\begin{equation}
\begin{split}
[(X,\Phi,\xi),(Y,\Psi,\eta)]_{E} =  \big( & [X,Y], \D_{X}\Psi - \D_{Y}\Phi - F(X,Y) - [\Phi,\Psi]_{\g}, \Li{X}\eta - i_{Y} \dr{\xi} \\
& -H(X,Y,\cdot) - \<F(\cdot,X),\Psi\>_{\g} + \<F(\cdot,Y), \Phi\>_{\g} + \< \D\Phi,\Phi'\>_{\g} \big),
\end{split}
\end{equation}
where $\D$ is the vector bundle connection induced by $A$ on $\g_{P}$, $F \in \Omega^{2}(M,\g_{P})$ is the curvature $2$-form of $A$, and $H \in \Omega^{3}(M)$ satisfies the condition 
\begin{equation}
\dr{H} + \frac{1}{2} \< F \^ F \>_{\g} = 0. 
\end{equation}
In particular, the first Pontriyagin class of $P$ must vanish. This structure is called a \textbf{heterotic Courant algebroid}, see \cite{bressler2007first} and \cite{Baraglia:2013wua}. 

Now, a generalized metric $V_{+}$ corresponds uniquely to the triple $(g,B,\vartheta)$, where $g$ is a Riemannian metric on $M$, $B \in \Omega^{2}(M)$ and $\vartheta \in \Omega^{1}(M,\g_{P})$. Note that this is true only for compact $G$, in general this is more complicated. Explicitly, one has 
\begin{equation}
\Gamma(V_{+}) = \{ (X, -\vartheta(X), (g + B - \frac{1}{2} \vartheta^{T} c_{\g} \vartheta)(X)) \; | \; X \in \X(M) \},
\end{equation}
where $c_{\g}: \g_{P} \rightarrow \g_{P}^{\ast}$ is the isomorphism induced by $\<\cdot,\cdot\>_{\g}$ and we view $\vartheta$ as a vector bundle map from $TM$ to $\g_{P}$. 

Writing $\omega$ in the same way as in example (1), condition (\ref{eq_divergence}) becomes 
\begin{equation}
\Div_{\cD}(X,\Phi,\xi) = \Div_{g}(X) - 2 (\dr{\phi})(X),
\end{equation}
for all $(X,\Phi,\xi) \in \Gamma(E)$. If we assume that $\cD \in \LC(E,V_{+},\Div_{\omega})$, the scalar curvature $\RS_{\cD}^{\gm}$ can be calculated and expressed in terms of $(g,B,\vartheta,\phi)$ to give
\begin{equation}
\RS_{\cD}^{\gm} = \RS(g) - \frac{1}{2} \<H',H'\>_{g} + \frac{1}{2} \dal F', F' \dar + 4 \Delta_{g}(\phi) - 4 \| \cD^{g}\phi \|^{2}_{g} + \frac{1}{6} \dim(\g),
\end{equation}
where the symbols have the same meaning as for type II supergravity example, except that
\begin{align}
H' = & \  H + \dr{B} - \frac{1}{2} \< \D \vartheta \^ \vartheta \>_{\g} - \frac{1}{6} \< [\vartheta \^ \vartheta]_{\g} \^ \vartheta \>_{\g} - \< F \^ \vartheta\>_{\g}, \\
F' = & \ F + \D{\vartheta} + \frac{1}{2} [\vartheta \^ \vartheta]_{\g},
\end{align}
where $\D{\vartheta} \in \Omega^{2}(M,\g_{P})$ is the exterior covariant derivative of $\vartheta$. By $\dal .,. \dar$, we denote the pairing on $\Omega^{1}(M,\g_{P})$ induced by the combination of $g$ and $\<\cdot,\cdot\>_{\g}$. Note that for the heterotic Courant algebroid, one has $\RS_{E} = - \frac{1}{6} \dim(\g)$. By plugging into the Einstein--Hilbert action, assuming $\partial M = \emptyset$ or $\dr{\phi}|_{\partial M} = 0$, we obtain
\begin{equation}
S_{EH}[g,B,\vartheta,\phi] = \int_{M} e^{-2\phi} \{ \RS(g) - \frac{1}{2} \<H',H'\>_{g} + \frac{1}{2} \dal F', F' \dar + 4 \| \cD^{g}\phi \|^{2}_{g} \} \cdot \omega_{g}. 
\end{equation}
For particular choices of $G$ and $P$, this action can be shown to be equivalent to the bosonic part of the heterotic supergravity \cite{polchinski2005string}. See \cite{Vysoky:2017epf} for detailed discussion and references. The equations of motion of this action form an example of the Strominger system, see \cite{garcia2016lectures}. 

\item \textbf{Symplectic gravity}: Consider the following general scenario first. 

Let $(E,\rho,\<\cdot,\cdot\>_{E},[\cdot,\cdot]_{E})$ be any Courant algebroid and suppose $L \subseteq E$ is its Dirac structure, that is a subbundle satisfying $L = L^{\perp}$ and involutive with respect to the bracket $[\cdot,\cdot]_{E}$. Note that this forces the signature $(p,q)$ of $\<\cdot,\cdot\>_{E}$ to satisfy $p = q = \rk(L)$. One calls $(E,L)$ a \textbf{Manin pair}. Next, one can always choose a Lagrangian subbundle $L' \subseteq E$, such that $E = L \oplus L'$. Note that choice is not unique and $L'$ is not necessarily involutive. We say that $(E,L,L')$ is a \textbf{split Manin pair}. For any Manin pair $(E,L)$, there is an induced Lie algebroid $(L,\ell,[\cdot,\cdot]_{L})$. 

Additionally, by fixing $L'$, we obtain the following data:
\begin{enumerate}[(i)]
\item A skew-symmetric $\R$-linear bracket $[\cdot,\cdot]_{L^{\ast}}$ on $\Gamma(L^{\ast})$ together with a vector bundle map $\ell^{\ast}: L^{\ast} \rightarrow TM$, satisfying the Leibniz rule, see Section 1-(1)-(a1). In general, this does \textit{not} make $(L^{\ast},\ell^{\ast},[\cdot,\cdot]_{L^{\ast}})$ into a Lie algebroid. 
\item A trivector $\dM \in \Gamma(\Lambda^{3}L)$. 
\end{enumerate}
Moreover, there are three non-trivial relations tying $(L,\ell,[\cdot,\cdot]_{L})$, $(L^{\ast},\ell^{\ast},[\cdot,\cdot]_{L^{\ast}})$ and $\dM$ together. This triple is called a \textbf{Lie quasi-bialgebroid}, see \cite{Bursztyn2007DiracGQ}.  

Finally, suppose $V_{+} \subseteq E$ is a generalized metric. It can be shown that it corresponds to a unique pair $(\mathbf{g}_{L},\Pi)$, where $\mathbf{g}_{L}$ is a fiber-wise metric on $L$ and $\Pi \in \Gamma(\Lambda^{2}L)$. 

Now, given $\cD \in \LC(E,V_{+})$ on a split Manin pair $(E,L,L')$, every $\cD$ can be uniquely described by a pair of tensors $W_{1,2} \in \Gamma(\Lambda^{2}L \otimes L)$. Its curvature tensor $R_{\cD}$ (and thus also $\Ric_{\cD}$ and $\RS_{\cD}^{\gm}$) can be then explicitly expressed in terms of $(L,L^{\ast},\dM)$, the generalized metric fields $(\mathbf{g}_{L},\Pi)$ and the fields $W_{1,2}$ describing the space of Levi-Civita connections. The calculation in fact only slightly generalizes the case of exact Courant algebroids described in Section 6 of \cite{Jurco:2016emw}. 

Let us consider the $E = \mathbb{T}M$ as in the example (1) above. Let $(g,B)$ be a pair describing the generalized metric $V_{+}$ and suppose $B$ is invertible. This allows one to consider a Dirac structure $L \subseteq E$ given by $\Gamma(L) := \{ (0,-B(X)) \; | \; B \in \X(M) \}$. The convenient choice of a complement $L'$ is then given by
\begin{equation}
\Gamma(L') := \{ (\theta(\xi), \xi) \; | \; \xi \in \Omega^{1}(M) \}, 
\end{equation}
where $\theta := B^{-1}$. Since we can obviously identify $L$ with $TM$, the split Manin pair $(E,L,L')$ induces the following data:
\begin{enumerate}[(i)]
\item The trivial Lie algebroid on $TM \cong L$.
\item The $\dr{B}$-twisted Koszul bracket on $\Omega^{1}(M) \cong \Gamma(L^{\ast})$ defined for all $\xi,\eta \in \Omega^{1}(M)$ by 
\begin{equation}
[\xi,\eta]_{\theta}^{\dr{B}} := \Li{\theta(\xi)}(\eta) - i_{\theta(\eta)} \dr{\xi} + \dr{B}(\theta(\xi),\theta(\eta),\cdot), 
\end{equation}
making $(T^{\ast}M, \theta, [\xi,\eta]_{\theta}^{\dr{B}})$ into a Lie algebroid. 
\item The trivector $\dM \in \X^{3}(M)$ given for all $\xi,\eta,\zeta \in \Omega^{1}(M)$ by 
\begin{equation}
\dM(\xi,\eta,\zeta) = -(H + \dr{B})(\theta(\xi),\theta(\eta),\theta(\zeta)). 
\end{equation}
\end{enumerate}
For the fields described by the generalized metric, one finds $G := \mathbf{g}_{L} = -Bg^{-1}B$ and $\Pi = 0$. Now, note that then $\omega_{g} = \omega_{G}^{\theta}$, where in any right-handed frame $(e_{i})_{i=1}^{n}$, one has 
\begin{equation}
\omega^{\theta}_{G} = \det(G)^{-\frac{1}{2}} |\det(\theta)|^{-1} e^{1} \^ \cdots \^ e^{n}. 
\end{equation}
Writing thus the general volume form $\omega$ as $\omega = e^{-2\phi} \omega^{\theta}_{G}$, for any $\cD \in \LC(E,V_{+},\Div_{\omega})$, one finds the following expression for the scalar curvatures:
\begin{equation}
\RS_{\cD}^{\gm} = \RS^{\theta}(G) + 4 \Div_{\theta}(\dr_{\theta} \phi) - 4 \| \dr_{\theta} \phi \|^{2}_{G} - \frac{1}{2} \< \dM, \dM\>_{G}, \; \; \RS_{\cD}^{E} = 0,
\end{equation}
where we use the following notation. $\RS^{\theta}(G)$ is the scalar curvature of the (unique) Levi-Civita connection on the Lie algebroid $(T^{\ast}M, \theta, [\cdot,\cdot]_{\theta}^{\dr{B}})$ with respect to the fiber-wise metric $G^{-1}$, $\Div_{\theta}$ is the induced divergence operator, and $\dr_{\theta}: C^{\infty}(M) \rightarrow \X^{1}(M)$ is the induced Lie algebroid differential. The Einstein--Hilbert action (\ref{eq_EHaction}) can be now written as a functional
\begin{equation} \label{eq_sympgravity}
S_{EH}[G,\theta,\phi] = \int_{M} e^{-2\phi} \{ \RS^{\theta}(G) + 4 \| \dr_{\theta}\phi \|^{2}_{G} - \frac{1}{2} \< \dM, \dM \>_{G} \} \cdot \omega_{G}^{\theta}. 
\end{equation}
Since we have started with the same Courant algebroid and a generalized metric $(g,B)$ as in the example (1) above, this can be viewed just as a redefinition of the fields of the type II supergravity. For $H = 0$, the action (\ref{eq_sympgravity}) was called a \textbf{symplectic gravity} in \cite{Blumenhagen:2012nt}. 
\item \textbf{Quadratic Lie algebra}: This is a rather trivial, yet important example. Suppose $M = \{ \ast \}$ is a singleton manifold. Any Courant algebroid over $M$ then reduces to a quadratic Lie algebra $(\d, \<\cdot,\cdot\>_{\d}, [\cdot,\cdot]_{\d})$, that is a Lie algebra with an invariant non-degenerate symmetric bilinear form.

Any volume form on $M$ is just a non-zero constant $\omega \neq 0$. Divergence operators are linear maps $\Div: \d \rightarrow \R$ and note that $\Div_{\omega} = 0$. A generalized metric is just a maximal positive definite subspace $V_{+} \subseteq \d$ with respect to $\<\cdot,\cdot\>_{\d}$. 

Observe that in this case, there is a canonical (but still not unique) connection $\cD \in \LC(\d,V_{+},0)$ defined for all $x,y,z \in \d$ by the formula 
\begin{equation} \label{eq_canconLA}
\begin{split}
\< \cD_{x}y, z \>_{\d} := & \ \frac{1}{3} \< [x_{+},y_{+}]_{\d}, z_{+}\>_{\d} + \frac{1}{3} \< [x_{-},y_{-}]_{\d},z_{-}\>_{\d} \\
& + \< [x_{+},y_{-}]_{\d}, z_{-}\>_{\d} + \< [x_{-},y_{+}]_{\d}, z_{+}\>_{\d},
\end{split}
\end{equation}
where $x_{\pm}$ are the components of $x \in \d$ with respect to the decomposition $\d = V_{+} \oplus V_{-}$. 

Note that it is easy to calculate the canonical scalar $\RS_{E}$. Since it can be calculated using any connection $\cD \in \LC(\d,V_{+},0)$ and \textit{any} generalized metric $V_{+}$, we may choose $V_{+} := \d$ (see also the remarks concluding Section \ref{sec_palatini}). The connection (\ref{eq_canconLA}) then takes the form $\cD_{x}y = \frac{1}{3}[x,y]_{\d}$. It is easy to see that 
\begin{equation}
R_{\cD}(x,y,z,w) = \frac{1}{6} \< [x,y]_{\d}, [z,w]_{\d} \>_{\d} 
\end{equation}
Consequently, one finds $\Ric_{\cD}(x,y) = -\frac{1}{6} c_{\d}(x,y)$, where $c_{\d}$ is the Cartan-Killing form of the Lie algebra $(\d,[\cdot,\cdot]_{\d})$. Note that $\Ric_{\cD}$ does not depend on $\<\cdot,\cdot\>_{\d}$. Hence
\begin{equation}
\RS_{E} = -\frac{1}{6} \Tr_{\mathbf{g}_{E}}(c_{\d}) \equiv \< \chi_{\d}, \chi_{\d} \>_{\d},
\end{equation}
where $\chi_{\d}(x,y,z) := \<[x,y]_{\d},z\>_{\d}$ is the canonical Cartan $3$-form associated to $(\d,\<\cdot,\cdot\>_{\d},[\cdot,\cdot]_{\d})$. Note that whenever $(\d,[\cdot,\cdot]_{\d})$ is a semi-simple Lie algebra, we may consider $\<\cdot,\cdot\>_{\d} := c_{\d}$ and the above formula gives $\RS_{E} = -\frac{1}{6} \dim(\d)$. 

Finally, the Einstein--Hilbert action in this example takes the form
\begin{equation}
S_{EH}[V_{+},\omega] := \{ \RS_{\cD}^{\gm} + \< \chi_{\d}, \chi_{\d} \>_{\d} \} \cdot \omega,
\end{equation}
and the corresponding equations of motion $\RS_{\cD}^{\gm} + \<\chi_{\d},\chi_{\d}\>_{\d} = 0$ and $\Ric_{\cD}(V_{+},V_{-}) = 0$ form a system of algebraic equations for the generalized metric $V_{+} \subseteq \d$ (they are independent of the volume form $\omega$). Finding solutions to these (very non-trivial) equations is of interest because of Poisson--Lie T-duality, see \cite{Jurco:2019tgt}. 
\end{enumerate}
\section{The proof} \label{sec_proof}
In this section, we will prove the Theorem \ref{thm_palatini}. We intend to only sketch the main steps and leave the details for the interested reader. Throughout this section, $(E,\rho,\<\cdot,\cdot\>_{E},[\cdot,\cdot]_{E})$ is a fixed but arbitrary Courant algebroid over an orientable manifold $M$.  Since we search for the extremals of the functional (\ref{eq_action}) of three independent variables $(V_{+},\omega,\cD)$, we will simply calculate the variations one by one. 
\begin{enumerate}[(1)]
\item \textbf{The generalized metric}: Suppose $V_{+} \subseteq E$ is a generalized metric. 

Any other generalized metric $V'_{+}$ can be always written as a graph $V'_{+} = \gr(\fvarphi_{+})$ of a unique vector bundle map $\fvarphi_{+}: V_{+} \rightarrow V_{-}$, that is 
\begin{equation}
\Gamma(\gr(\fvarphi_{+})) = \{ \psi_{+} + \fvarphi_{+}(\psi_{+}) \; | \; \psi_{+} \in \Gamma(V_{+}) \}.
\end{equation}
We can use this observation to consider an \textit{arbitrary} compactly supported vector bundle map $\fvarphi_{+}: V_{+} \rightarrow V_{-}$ vanishing on $\partial M$. For each $\epsilon > 0$, define a new subbundle $V'_{+}(\epsilon) := \gr(\epsilon \varphi_{+})$. For $\epsilon$ sufficiently small, this defines a new generalized metric on $(E,\<\cdot,\cdot\>_{E})$. Let $\gm'(\epsilon)$ be the corresponding fiber-wise metric on $E$. 

Now, let $\A: \Gamma(E) \times \Gamma(E) \rightarrow C^{\infty}(M)$ be an arbitrary \textit{symmetric} $C^{\infty}(M)$-bilinear map. A straightforward calculation leads to the relation
\begin{equation}
\Tr_{\gm'(\epsilon)}(\A) = \Tr_{\gm}(\A) + 4 \epsilon \cdot \Tr_{\mathbf{g}_{+}}( \A_{+-} \circ (\1_{V_{+}} \times \fvarphi_{+})) + O(\epsilon^{2}),
\end{equation}
where $\A_{+-}$ denotes the restriction of $\A$ to $\Gamma(V_{+}) \times \Gamma(V_{-})$ and $\mathbf{g}_{+}$ is the positive definite metric induced on $V_{+}$ by the restriction of $\<\cdot,\cdot\>_{E}$. If $\{ \Phi^{+}_{\lambda} \}_{\lambda=1}^{p}$ and $\{ \Phi^{-}_{\mu} \}_{\mu=1}^{q}$ are local frames for $V_{+}$ and $V_{-}$, respectively, one can write the  term linear in $\epsilon$ locally as 
\begin{equation}
4 \epsilon \cdot \Tr_{\gm_{+}}( \A_{+-} \circ (\1_{V_{+}} \times \fvarphi_{+})) = 4 \epsilon \cdot \A_{+-}( \Phi^{+}_{\lambda}, \Phi^{-}_{\mu} ) \cdot [\fvarphi_{+}]^{\mu \lambda},
\end{equation}
where $[\fvarphi_{+}]^{\mu \lambda} := \Phi^{\mu}_{-}\{ \fvarphi_{+}(\mathbf{g}_{+}^{-1}(\Phi^{\lambda}_{+}))\}$ with $\{ \Phi^{\lambda}_{+} \}_{\lambda=1}^{p}$ and $\{ \Phi^{\mu}_{-} \}_{\mu=1}^{q}$ being local frames dual to the above ones. Plugging these expressions into the Palatini action (\ref{eq_action}), one gets
\begin{equation}
S[V'_{+}(\epsilon),\omega,\cD] = S[V_{+},\omega,\cD] + 4 \epsilon \cdot \int_{M} [\Ric_{\cD}]_{+-}(\Phi^{+}_{\lambda}, \Phi^{-}_{\mu}) \cdot [\fvarphi_{+}]^{\mu \lambda} \cdot \omega + O(\epsilon^{2}). 
\end{equation}
Since $[\fvarphi_{+}]^{\mu \lambda}$ are arbitrary, we conclude that $V_{+}$ is an extremal generalized metric, iff
\begin{equation}
\Ric_{\cD}(\Gamma(V_{+}),\Gamma(V_{-})) = 0,
\end{equation} 
that is $\cD$ is Ricci compatible with $V_{+}$, as claimed by Theorem \ref{thm_palatini}-$(i)$. 
\item \textbf{The volume form}: This one is rather easy. Let $\omega$ be a given volume form. Then one can consider a variation $\omega'(\epsilon) := e^{\epsilon \phi} \cdot \omega$, where $\phi \in C^{\infty}(M)$ is an arbitrary compactly supported smooth function vanishing on $\partial M$. It follows that 
\begin{equation}
S[V_{+},\omega'(\epsilon), \cD] = S[V_{+},\omega,\cD] + \epsilon \cdot \int_{M} \{ \RS_{\cD}^{\gm} + \RS_{E} \} \cdot \phi \cdot \omega + O(\epsilon^{2}). 
\end{equation} 
Since $\phi$ is arbitrary, this proves that $\omega$ is an extremal volume form, if and only if 
\begin{equation}
\RS_{\cD}^{\gm} + \RS_{E} = 0. 
\end{equation}
This corresponds to the statement $(ii)$ of Theorem \ref{thm_palatini}, as was to be proved. 
\item \textbf{The connection}: Let $\cD$ be any Courant algebroid connection. The most general variation of the connection takes the form 
\begin{equation}
\<[\cD'(\epsilon)](\psi,\psi'),\psi''\>_{E} := \<\cD(\psi,\psi'),\psi''\>_{E} + \epsilon \cdot \cN(\psi,\psi',\psi''),
\end{equation}
where $\cN \in \Gamma(E^{\ast} \otimes \Lambda^{2}E^{\ast})$ is arbitrary and vanishing on $\partial M$. To find the conditions on $\cD$, let us write it as follows. Suppose $\cD^{0} \in \LC(E,V_{+},\Div_{\omega})$ be an auxiliary Levi-Civita connection with its divergence operator set to $\Div_{\omega}$. There is thus a unique $\K \in \Gamma(E^{\ast} \otimes \Lambda^{2}E^{\ast})$ such that 
\begin{equation} \label{eq_cDusingcD0K}
\<\cD(\psi,\psi'),\psi''\>_{E} = \< \cD^{0}(\psi,\psi'), \psi''\>_{E} + \K(\psi,\psi',\psi'').
\end{equation}
The properties of $\cD$ can be now fully encoded by the conditions imposed on $\K$ and the above variation can be viewed as considering a connection parametrized by $\cD^{0}$ and a new tensor field $\K'(\epsilon) = \K + \epsilon \cN$, instead. Note this is an idea similar to \cite{dadhich2012equivalence}. Before the actual calculation, let us introduce some notation. 

First, if $\{ \psi_{\lambda} \}_{\lambda=1}^{\rk(E)}$ is some local frame for $E$, by $\{ \psi^{\lambda} \}_{\lambda=1}^{\rk(E)}$ we denote the dual local frame. By $\{ \psi^{\lambda}_{E} \}_{\lambda=1}^{\rk(E)}$ and $\{ \psi^{\lambda}_{\gm} \}_{\lambda=1}^{\rk(E)}$ we denote the local frames for $E$ obtained from the dual frame using $\<\cdot,\cdot\>_{E}$ and $\gm$, respectively. Next, let us consider two sections $\K'$ and $\K'_{\gm}$ of $E^{\ast}$ defined as ``partial traces'' of $\K$, that is for all $\psi \in \Gamma(E)$, set
\begin{equation} \label{eq_partialtraces}
\K'(\psi) := \K(\psi_{\lambda},\psi^{\lambda}_{E}, \psi), \; \; \K'_{\gm}(\psi) := \K(\psi_{\lambda},\psi^{\lambda}_{\gm},\psi). 
\end{equation}
The quantities $\cN'$ and $\cN'_{\gm}$ are defined in the same way. We claim (without a proof) that there holds the relation
\begin{equation}
\begin{split}
\RS_{\cD'(\epsilon)}^{\gm} = \RS_{\cD}^{\gm} + \epsilon \cdot \{&\Div_{\omega}(\cN'_{\gm} \circ \mathbf{g}_{E}^{-1}) + \Div_{\omega}( \cN' \circ \gm^{-1}) \\
& + \frac{1}{2} \V(\psi_{\lambda}, \psi_{\mu},\psi_{\nu}) \cdot \cN(\psi^{\lambda}_{E}, \psi^{\mu}_{E}, \psi^{\nu}_{E}) \} + O(\epsilon^{2}),
\end{split}
\end{equation}
where $\V \in \Gamma(E^{\ast} \otimes \Lambda^{2}E^{\ast})$ depends only on $\K$ and $V_{+}$ and takes the form 
\begin{equation}
\begin{split}
\V(\psi,\psi',\psi'') = & \ \< \psi, \tau(\psi'')\>_{E} \cdot \K'(\psi') - \< \psi, \tau(\psi')\>_{E} \cdot \K'(\psi'') \\
& + \< \psi, \psi''\>_{E} \cdot \K'_{\gm}(\psi') - \< \psi,\psi'\>_{E} \cdot \K'_{\gm}(\psi'') \\
& + \K(\psi,\tau(\psi''),\psi') - \K(\psi,\tau(\psi'),\psi'') + \K(\tau(\psi'),\psi,\psi'') \\
& - \K(\tau(\psi''),\psi,\psi') + \K(\psi',\tau(\psi),\psi'') - \K(\psi'',\tau(\psi),\psi'),
\end{split}
\end{equation}
for all $\psi,\psi',\psi'' \in \Gamma(E)$. Next, note that for every section $\psi \in \Gamma(E)$ satisfying $\psi|_{\partial M} = 0$, one has $\int_{M} \Div_{\omega}(\psi) \cdot \omega = 0$. Since this is true for both $\cN'_{\gm} \circ \mathbf{g}_{E}^{-1}$ and $\cN' \circ \gm^{-1}$, we find 
\begin{equation}
S_{P}[V_{+},\omega,\cD'(\epsilon)] := S_{P}[V_{+},\omega,\cD] + \frac{\epsilon}{2} \int_{M} \V(\psi_{\lambda},\psi_{\mu},\psi_{\nu}) \cdot \cN(\psi^{\lambda}_{E},\psi^{\mu}_{E},\psi^{\nu}_{E}) \cdot \omega + O(\epsilon^{2}). 
\end{equation}
We see that $\cD$ is an extremal of $S_{P}$, iff $\V = 0$. We claim that this is equivalent to $\cD \in \LC(E,V_{+},\Div_{\omega})$. Recall that we have parametrized $\cD$ as in (\ref{eq_cDusingcD0K}). Let us now rephrase the required properties in terms of the tensor field $\K$:
\begin{enumerate}[(a)]
\item $\cD$ is compatible with the generalized metric $V_{+}$, iff $\K(\psi,\psi'_{+},\psi''_{-}) = 0$ for all $\psi \in \Gamma(E)$, $\psi'_{+} \in \Gamma(V_{+})$ and $\psi''_{-} \in \Gamma(V_{-})$.
\item $\cD$ is torsion-free, iff the complete skew-symmetrization $\K_{a}$ of $\K$ vanishes.
\item $\Div_{\cD} = \Div_{\omega}$, iff $\K' = 0$. 
\end{enumerate}
Let us now argue that $\V = 0$ is equivalent to the all three conditions on $\K$ in (a) - (c). In other words, we prove that $\V = 0$, iff $\cD \in \LC(E,V_{+},\Div_{\omega})$, the statement $(iii)$ of Theorem \ref{thm_palatini}. Consequently, this will conclude the proof.

Fist, assume that $\V = 0$. Consequently, the two partial traces $\V', \V'_{\gm} \in \Gamma(E^{\ast})$ must also vanish, see (\ref{eq_partialtraces}). For each $\psi \in \Gamma(E)$, one finds the system of equations 
\begin{align}
\V'(\psi) = & \ - \Tr(\tau) \cdot \K'(\psi) + (2 - \rk(E)) \cdot \K'_{\gm}(\psi) = 0, \\
\V'_{\gm}(\psi) = & \ (2 - \rk(E)) \cdot \K'(\psi) - \Tr(\tau) \cdot \K'_{\gm}(\psi) = 0.
\end{align}
We solve for $\K'(\psi)$ and $\K'_{\gm}(\psi)$. If  $(p,q)$ is the signature of $\<\cdot,\cdot\>_{E}$, we have $\rk(E) = p + q$ and $\Tr(\tau) = p - q$. The matrix of this system is singular, iff $\Tr(\tau) = \pm (2 - \rk(E))$, that is $p = 1$ or $q = 1$. The assumption $p,q \neq 1$ made just above Theorem \ref{thm_palatini} thus forces $\K' = \K'_{\gm} = 0$. In particular, we have just proved that $\V = 0$ implies the condition (c). 

With this in mind, let us evaluate $\V = 0$ on a various combinations of sections, where $\pm$ always indicates an element of $\Gamma(V_{\pm})$. One finds
\begin{align}
\label{eq_Vmixed1} \V(\psi_{+},\psi'_{+},\psi''_{-}) = & \  2 \K(\psi'_{+},\psi_{+},\psi''_{-}) = 0, \\ 
\label{eq_Vmixed2} \V(\psi_{-},\psi'_{+},\psi''_{-}) = & \ -2 \K(\psi''_{-},\psi'_{+},\psi_{-}) = 0, \\
\label{eq_Vmixed3} \V(\psi_{+},\psi'_{-},\psi''_{-}) = & \ 2 \K(\psi_{+},\psi'_{-},\psi''_{-}) = 0, \\
\label{eq_Vmixed4} \V(\psi_{-},\psi'_{+},\psi''_{+}) = & \ -2 \K(\psi_{-},\psi'_{+},\psi''_{+}) = 0. 
\end{align}
This proves that all ``mixed'' components of $\K$ have to vanish. In particular, this forces the condition in (a) to hold. Finally, one obtains 
\begin{align}
\label{eq_Vhomogeneous1} \V(\psi_{+},\psi'_{+},\psi''_{+}) = & \ - 6 \K_{a}(\psi_{+},\psi'_{+},\psi''_{+}) = 0, \\
\label{eq_Vhomogeneous2} \V(\psi_{-},\psi'_{-},\psi''_{-}) = & \ 6 \K_{a}(\psi_{-},\psi'_{-},\psi''_{-}) = 0. 
\end{align}
We see that due to (\ref{eq_Vmixed1} - \ref{eq_Vmixed4}), this implies $\K_{a} = 0$, that is the torsion-free condition (b). We conclude that $\V = 0$ indeed implies $\cD \in \LC(E,V_{+},\Div_{\omega})$. 

Conversely, assuming that $\cD \in \LC(E,V_{+},\Div_{\omega})$, we have $\K' = 0$ by (c). The compatibility with $V_{+}$ implies that that $\K'_{\gm}(\psi) = \K'(\tau(\psi))$ for all $\psi \in \Gamma(E)$ and thus also $\K'_{\gm} = 0$. The metric compatibility (a) implies (\ref{eq_Vmixed1}, \ref{eq_Vmixed2}). The torsion-free condition (c) gives (\ref{eq_Vhomogeneous1}, \ref{eq_Vhomogeneous2}) and together with (a) also (\ref{eq_Vmixed3}, \ref{eq_Vmixed4}). This exhausts all $\pm$ possibilities and we conclude that $\V = 0$. 
\end{enumerate}
\section{Conclusion and Outlook}
We believe that the results presented in this paper are an interesting contribution to the understanding of geometric foundations of string theory at its low-energy effective action.
They extend the Palatini formalism to the generalized Riemannian geometry of Courant algebroids and lead to the correct low-energy effective actions. What we found especially intriguing is the fact that in the framework of generalized geometry used here, we can start from an arbitrary Courant algebroid connection. Then the Levi-Civita property - including, in contrary to the ordinary case also the torsionless condition - follows from equations of motion (Palatini variation). In addition, Palatini variation forces the divergence of Levi-Civita connection to be compatible with the volume form (dilaton). Although the set of Levi-Civita connections with fixed divergence is infinite, the resulting equations of motion and the effective action doesn't depend on the choice of such a connection. All this can be understood as a posteriori justification of the generalized geometry notions of connection, torsion and curvature introduced in the earlier literature cited in the previous sections. Also, it suggests that the dilaton can be used, at least partially, to restrict the set of physically relevant Levi-Civita connections.

What we didn't discuss in the paper, is 
a) compatibility of the Palatini variation with reductions of Courant algebroids (Kaluza--Klein type of reductions) and, hence, with $T$-duality and b) a proper modification of Palatini formalism in the context of double field theory. We hope return to these questions in forthcoming papers.
Also, we hope to extend our approach, in order to include R-R fields and fermions, in the future.

\bibliography{bib}
\appendix

\end{document}